\documentclass[twocolumn,twoside,slac_two]{revtex4}
\usepackage{graphicx}
\usepackage{fancyhdr}
\begin{document}
\title{New Detectors for the Kaon and Hypernuclear Experiments with
KaoS at MAMI and with PANDA at GSI}

\author{P. Achenbach}
\email[E-mail address: ]{patrick@kph.uni-mainz.de}
\author{C. \surname{Ayerbe Gayoso}}
\author{R. B{\"o}hm}
\author{M.O. Distler}
\author{J. Friedrich}
\author{K.W. Krygier}
\author{H. Merkel}
\author{U. M{\"u}ller}
\author{R. Neuhausen}
\author{L. Nungesser}
\author{J. Pochodzalla}
\author{A. \surname{Sanchez Lorente}}
\author{S. \surname{S{\'a}nchez Majos}}
\author{Th. Walcher}
\affiliation{Institut f{\"u}r Kernphysik, Johannes Gutenberg-Universit{\"a}t
    Mainz, Germany}

\author{J. Gerl}
\author{M. Kavatsyuk}
\author{I. Kojouhavorv}
\author{N. Saito}
\author{T.R. Saito}
\author{H. Schaffner}
\affiliation{GSI, Darmstadt, Germany}

\author{T. Bressani}
\author{S. Bufalino}
\author{A. Feliciello}
\affiliation{Dipartimento di Fisica Sperimentale, Universit{\`a} di Torino and
    INFN Sezione di Torino, Italy} 

\author{A. Pantaleo}
\author{M. Palomba}
\affiliation{Dipartimento di Fisica, Universit{\`a} di Bari, and
        INFN Sezione di Bari, Italy} 

\author{G. Raciti}
\author{C. Sfienti}
\affiliation{Dipartimento di Fisica, Universit{\`a} di Catania, and
    INFN Sezione di Catania, Italy} 

\author{M. Agnello}
\author{F. Ferro}
\author{F. Iazzi}
\author{K. Szymanska}
\affiliation{Dipartimento di Fisica del Politecnico di Torino,
    Italy} 

\author{P.-E. Tegn{\'e}r}
\affiliation{Department of Physics, Stockholm University, Sweden}

\author{B. Cederwall}
\affiliation{Department of Physics, Royal Institute of Technology, Stockholm,
Sweden}

\author{L. Majling}
\affiliation{Nuclear Physics Institute, Academy of Sciences of the Czech
    Republic, Rez near Prague, Czech Republic}

\author{(A1 Collaboration and HyperGamma Collaboration)}\noaffiliation

\begin{abstract}
  The KaoS spectrometer at the Mainz Microtron MAMI, Germany, is
  perceived as the ideal candidate for a dedicated spectrometer in
  kaon and hypernuclei electroproduction. KaoS will be equipped with
  new read-out electronics, a completely new focal plane detector
  package consisting of scintillating fibres, and a new trigger
  system. First prototypes of the fibre detectors and the associated
  new front-end electronics are shown in this contribution.  The Mainz
  hypernuclei research program will complement the hypernuclear
  experiments at the planned FAIR facility at GSI, Germany.  At the
  proposed antiproton storage ring the spectroscopy of double
  $\Lambda$ hypernuclei is one of the four main topics which will be
  addressed by the $\overline{\mbox{\sf P}}${\sf ANDA}\
  Collaboration. The experiments require the operation of high purity
  germanium (HPGe) detectors in high magnetic fields ($B\approx 1\,$T)
  in the presence of a large hadronic background.  The performance of
  high resolution Ge detectors in such an environment has been
  investigated.
\end{abstract}

\maketitle

\thispagestyle{fancy}
\section{Strange Hadrons with KaoS at MAMI}
Strangeness production in the energy regime $1-2$\,GeV is undergoing a
renewed interest, both theoretically and experimentally. On the
theoretical side, due to the non-perturbative nature of Quantum
Chromodynamics (QCD) at low energies, kaon electroproduction cannot be
described by the fundamental equations for the dynamics of
(asymptotically free) quarks and gluons.  Instead, isobaric models are
commonly used, where the hadrons are treated as effective degrees of
freedom. The partonic constituents can also be considered along the
lines of chiral models, which take an important feature of low energy
QCD into account, namely the chiral symmetry and its spontaneous
breakdown. Although lattice QCD calculations are not yet relevant to
kaon production, it is anticipated that precise experimental data on
strangeness production will challenge and improve our understanding of
the strong interaction in the low energy regime of QCD.

Experimentally, the conservation of strangeness in electromagnetic and
strong interactions allows the tagging of baryonic systems with open
strangeness, e.g.\ baryon resonances or hypernuclei, by detecting a
kaon in the final channel. Currently, only few high quality data
points exist on $K^+\Lambda$ electroproduction close to threshold, the
available information on $K^+\Sigma^0$ is even more sparse.  This
field of physics will be addressed at Mainz Microtron MAMI, Germany,
with the KaoS spectrometer.

KaoS is a very compact magnetic spectrometer suitable especially for
the detection of kaons. It was built for the GSI in Darmstadt,
Germany, for heavy ion induced experiments~\cite{Senger1993}. During
May and June 2003 the KaoS magnets together with associated
electronics and detectors were brought to Mainz. As a pilot experiment
on kaon electroproduction, the separation of transverse and
longitudinal structure functions in parallel kinematics is
planned~\cite{PAC2003}. For the pilot experiment only the detection of
positive kaons at moderate angles, $\theta \sim 10^\circ$, is
required. With the available detector package charged particle
trajectories can be measured by using two multi-wire chambers of
120\,cm length. Furthermore, time-of-flight (TOF) and trigger
information can be obtained from a segmented scintillator array.  The
use of KaoS as a two arm spectrometer for the electroproduction of
hypernuclei at MAMI requires the detection of the scattered electron
at laboratory angles close to 0$^{\circ}$ with typical momenta of
650\,MeV$/c$.  The kaon detector in the focal plane has to cover a
range of scattering angles around 5$^{\circ}$ in coincidence.  In
order to cope with the special kinematics for electroproduction of
hypernuclei and with the high rates background rates, KaoS will be
equipped with new read-out electronics, a completely new focal plane
detector package for the electron arm, and a new trigger system.

\subsection{Development of new Focal Plane Detectors}
\begin{figure}[htb]
  \includegraphics[width=\columnwidth]{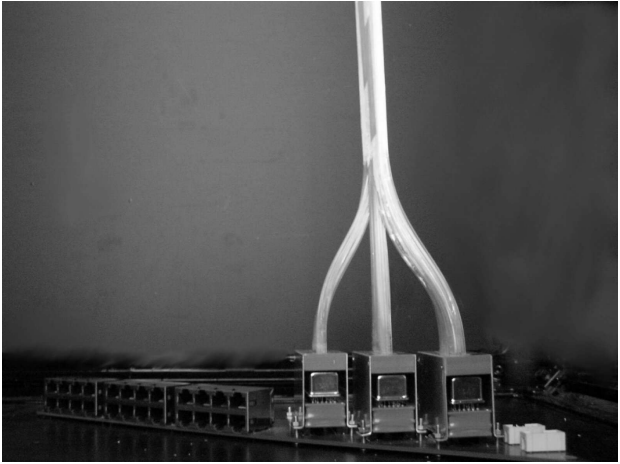}
  \caption{Photograph of a prototype triple detector with three
  joined fibre bundles of 58.4\,mm width, bare MAPMTs, and 
  Cockcroft-Walton voltage multipliers mounted on a single 
  front-end board.}
  \label{fig:triple_detector}
\end{figure}

The main focal plane detector of the KaoS electron arm will consist of
2 horizontal planes of fibre arrays, covering an active area of $1500
\times 500$\,mm$^2$, and comprising close to 2000 channels per plane
(63 detectors on 21 triple boards).  Each plane is divided into
58.4\,mm wide triple detectors which consist of 384 fibres in three
joined fibre segments coupled to three multi-anode photomultipliers
(MAPMTs) and Cockcroft-Walton voltage multipliers mounted on a single
96-channel front-end board. A prototype triple detector was built and
tested, see Fig.~\ref{fig:triple_detector} for a photograph.

A fibre doublet structure is formed from two single layers of fibres,
with one of the fibre layers off-set relative to the other by half a
fibre spacing. The virtue of this configuration is the high fraction
of overlapping fibres, i.e.\ a high detection efficiency, and a small
pitch leading to a good spatial resolution.  Several of such double
layers are introduced if the number of photoelectrons per fibre is too
small to be detected with the required efficiency.

The fibres are of type {\sf Kuraray} SCSF-78 with double cladding and
of $0.83$\,mm outer diameter.  The cladding thickness is $0.1\,$mm,
leading to a 0.73\,mm core of refractive index $n_{\it core}=$
1.6. Four fibres are grouped to one channel and brought to one pixel
of the MAPMT.

\begin{figure}[htb]
  \includegraphics[width=0.42\columnwidth]{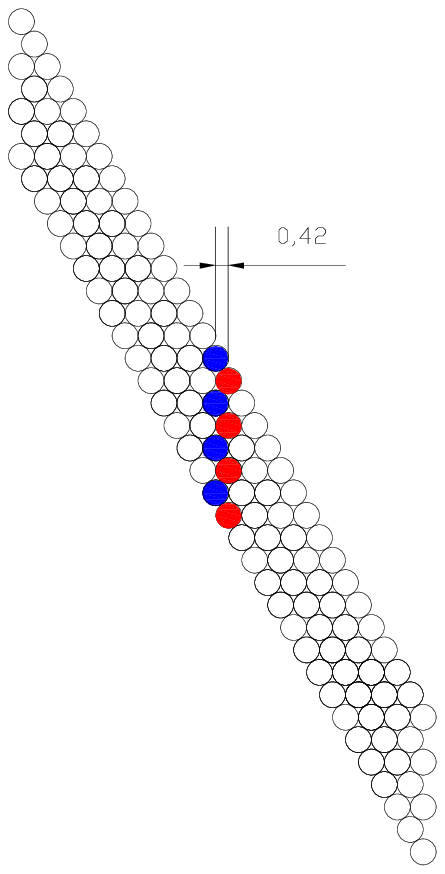}\hfill
  \includegraphics[angle=90,width=0.35\columnwidth]{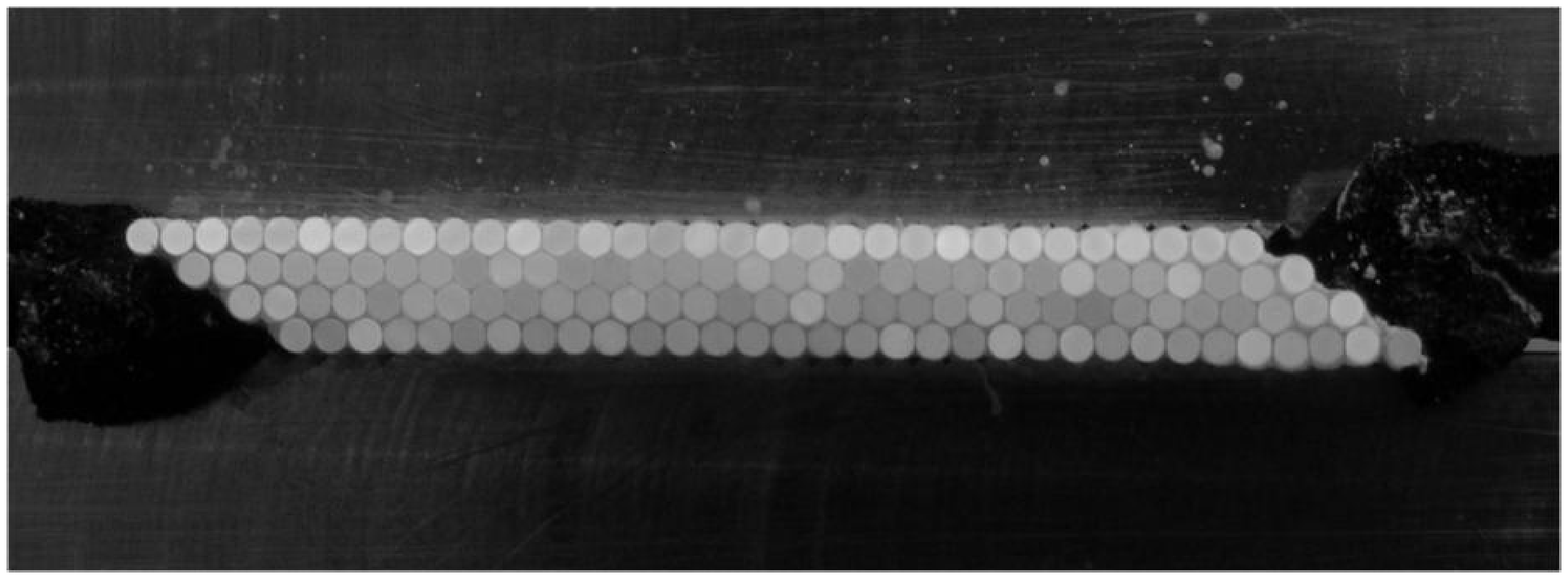}
  \caption{Scheme of a 60$^\circ$ column angle configuration with 4
  double layers and a pitch of 0.42\,mm (left) and a photograph of an
  assembled 60$^\circ$ fibre bundle (right).}
  \label{fig:60deg}
\end{figure}

The MAPMTs of type R7259K from {\sf Hamamatsu} are based on 32-channel
linear array multi-anodes.  The tubes have been delivered without
base. The photocathode material is bialkali and the window is made of
borosilicate glass. The effective area per channel is $0.8 \times
7$\,mm$^2$ with a pitch of 1\,mm.  Our tubes have an average anode
luminous sensitivity of $S_a= 374\,$A$/$lm (according to data sheet
140\,A$/$lm is typical), an average cathode luminous sensitivity
$S_k=84\,\mu$A$/$lm (70\,$\mu$A$/$lm typical), and an average gain $G=
4.4 \cdot 10^6$ ($2\cdot 10^6$ typical). The gain uniformity between
anodes lies between 1:1.1 and 1:1.25 (1:1.5 typical), with the edge
anodes having slightly lower gains on average. Hardly any strip has
less than 70\% of the maximum gain in a given array. The photocathode
sensitivity is defined as the ratio of the cathode current $I_k$ (less
the dark current) to the incident flux $\Phi$, expressed in
photometric units: $S_k(\mathrm{A}/\mathrm{lm})=
\frac{I_k(\mathrm{A})}{\Phi(\mathrm{lm})}$.

Instead of supplying dynode voltages through a voltage divider, the
phototubes are powered by individual Cockcroft-Walton bases,
manufactured by {\sf HVSys}, Dubna. The dc voltage is pulsed and
converted with a voltage doubler ladder network of capacitors and
diodes to higher voltages. The principal advantage is that there is no
need for stiff high voltage cables, since only $\sim 140$\,V has to be
distributed to the first front-end board, where the voltage is
daisy-chained to the other boards of the detector plane. one drawback
is that the voltages can only be equally spaced, which is acceptable
for their actual use.

The first prototypes were designed for 128 fibres packed in 4 double
layers in 0$^\circ$ column angle configuration, with a pitch of
0.6\,mm between adjacent columns. Such a geometry implies an incident
angle for scattered electrons of 0$^\circ$.  For a set-up in the
electron focal plane of the KaoS spectrometer, this column angle is
not appropriate since the scattered electrons have an inclination
angle of $50-70^\circ$ with respect to the normal of the focal plane.
Instead, a configuration with a column angle of $\alpha= 60^\circ$ and
hexagonal packing has been chosen. The hexagonal packing, in which the
centres of the fibres are arranged in a hexagonal lattice, and each
fibre is surrounded by 6 other fibres, has the highest density of
$\frac{\pi}{\sqrt{12}}\simeq 0.9069$. A prototype triple detector with
this configuration is shown in Fig.~\ref{fig:60deg}. The small space
between adjacent triple boards, with a pitch of only 79.6\,mm, poses
some difficulties on positioning and alignment. The small dimension of
the dynode channels on the electrode plate, $0.8$\,mm, and the
diameter of the fibres, $0.83$\,mm, makes it clear that the alignment
has to be very precise.

\begin{figure}[htb]
  \includegraphics[width=\columnwidth]{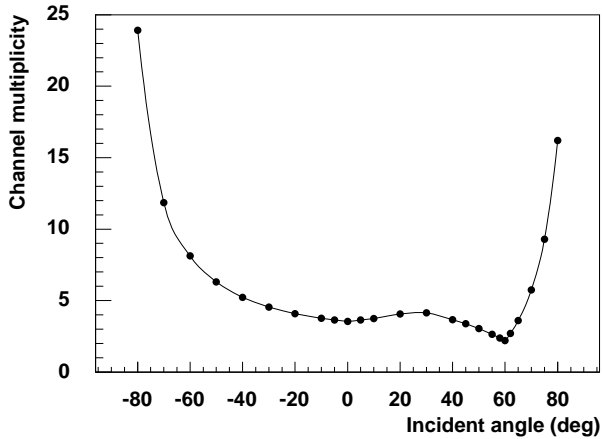}
  \caption{Simulated variation of the channel multiplicity as a
  function of the incident angle for a fibre array configuration with
  60$^\circ$ column angle. The incident angles in the
  electron focal plane  are distributed between $50^\circ-70^\circ$.}
  \label{fig:multiplicities}
\end{figure}

Multiple scattering through small angles is given by the Gaussian
width $\theta_0= \theta^{RMS}_{plane}=
13.6\,\mbox{MeV}\frac{(z=1)}{(\beta = 1)cp} \cdot \sqrt{x/X_0} \cdot
\left(1 + 0.038\,\ln{x/X_0}\right)$~\cite{Bethe1953,Scott1963}.  The
amount of scattering depends primarily on the momentum and the
thickness of the scattering medium (radiation length $X_0= 42.4\,$cm
for a polystyrene scintillator). In the focal plane detector the
latter depends also slightly on the momentum because tracks for
different momenta pass the focal plane detector at different places
under slightly different angles.  The detector configuration leads to
a thickness variation of $\pm 40\%\approx \pm 0.3\,$mm$/$layer.  The
average thickness and its variation was simulated for electrons
traversing the focal plane to be $x= (4.70 \pm 1.29)$\,mm.  This
number translates into a width of $\theta_0= 0.227^\circ$ for $p=
300$\,MeV$/c$.  The main consequence of the incident angle
distribution is an increased channel multiplicity. The pure
geometrical effect was studied in a simulation of a $60^\circ$
detector, the full calculation is shown in
Fig.~\ref{fig:multiplicities}. The incident angles in the electron
focal plane are distributed between $50^\circ-70^\circ$.

\subsection{Development of new Read-Out Electronics}
\begin{figure}[htb]
  \includegraphics[width=\columnwidth]{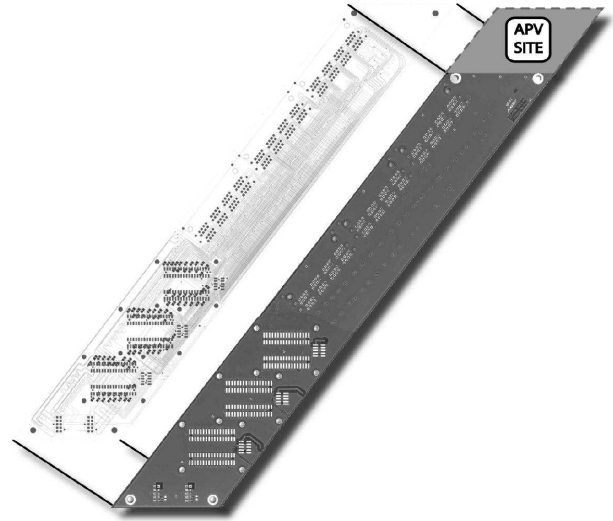}
  \caption{Photograph and circuit scheme of the triple front-end
  board showing the three PMT sockets at the lower left, and the
  output sockets. The site for the APV chip connector is indicated.}
  \label{fig:fe-board}
\end{figure}

A 12-layer front-end board able to accommodate three 32-channel
multi-anode photomultipliers with minimum time jitter was designed,
see Fig.~\ref{fig:fe-board}. It houses the low voltage power supply
for the Cockcroft-Walton voltage multiplier bases, the RJ-45
connectors for analogue output to the discriminators and, in the near
future, APV25 chips for amplifying, sampling and multiplexing the
signal amplitudes.

\begin{figure}[htb]
  \includegraphics[width=0.50\columnwidth]{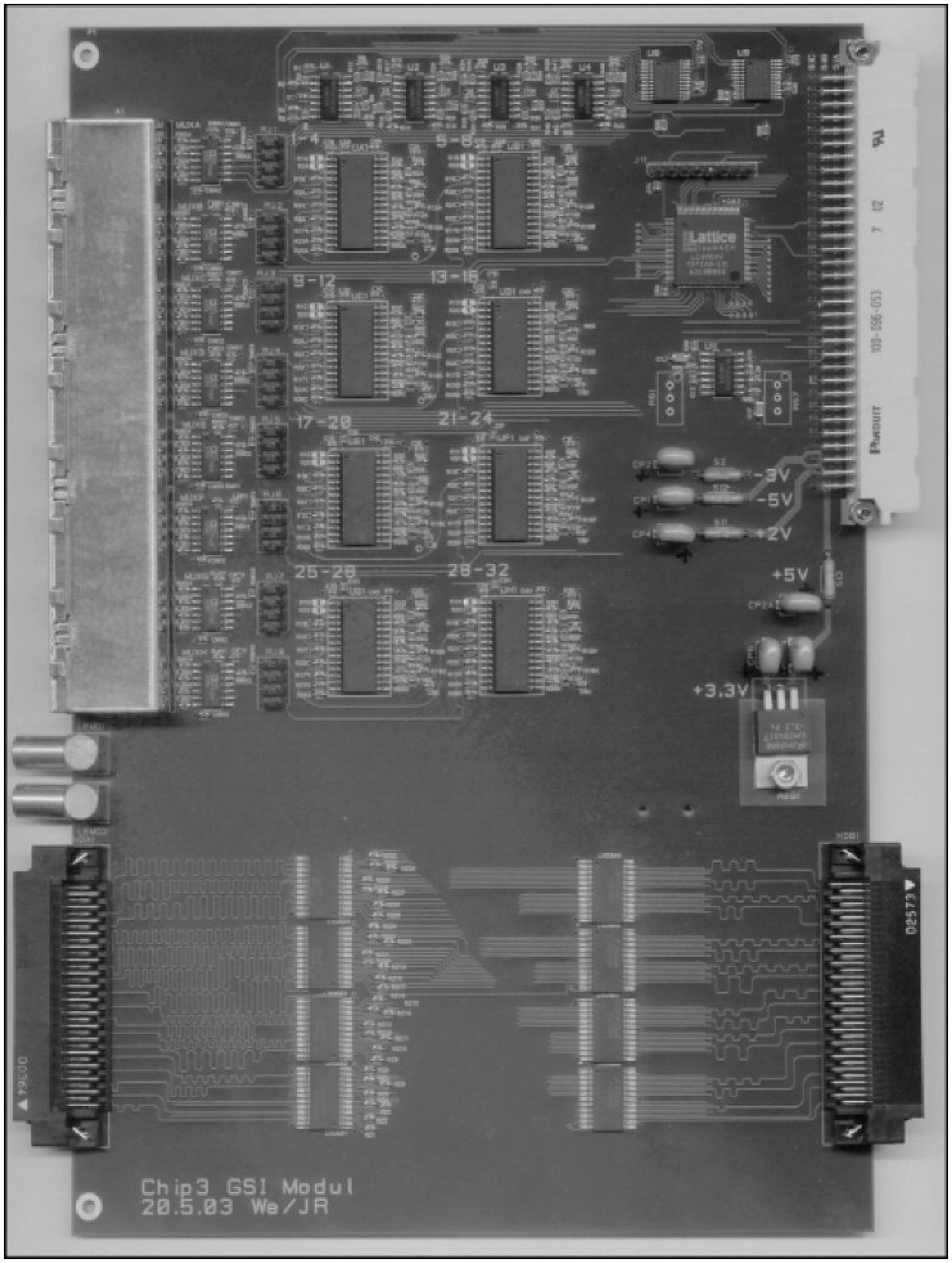}
  \includegraphics[width=0.475\columnwidth]{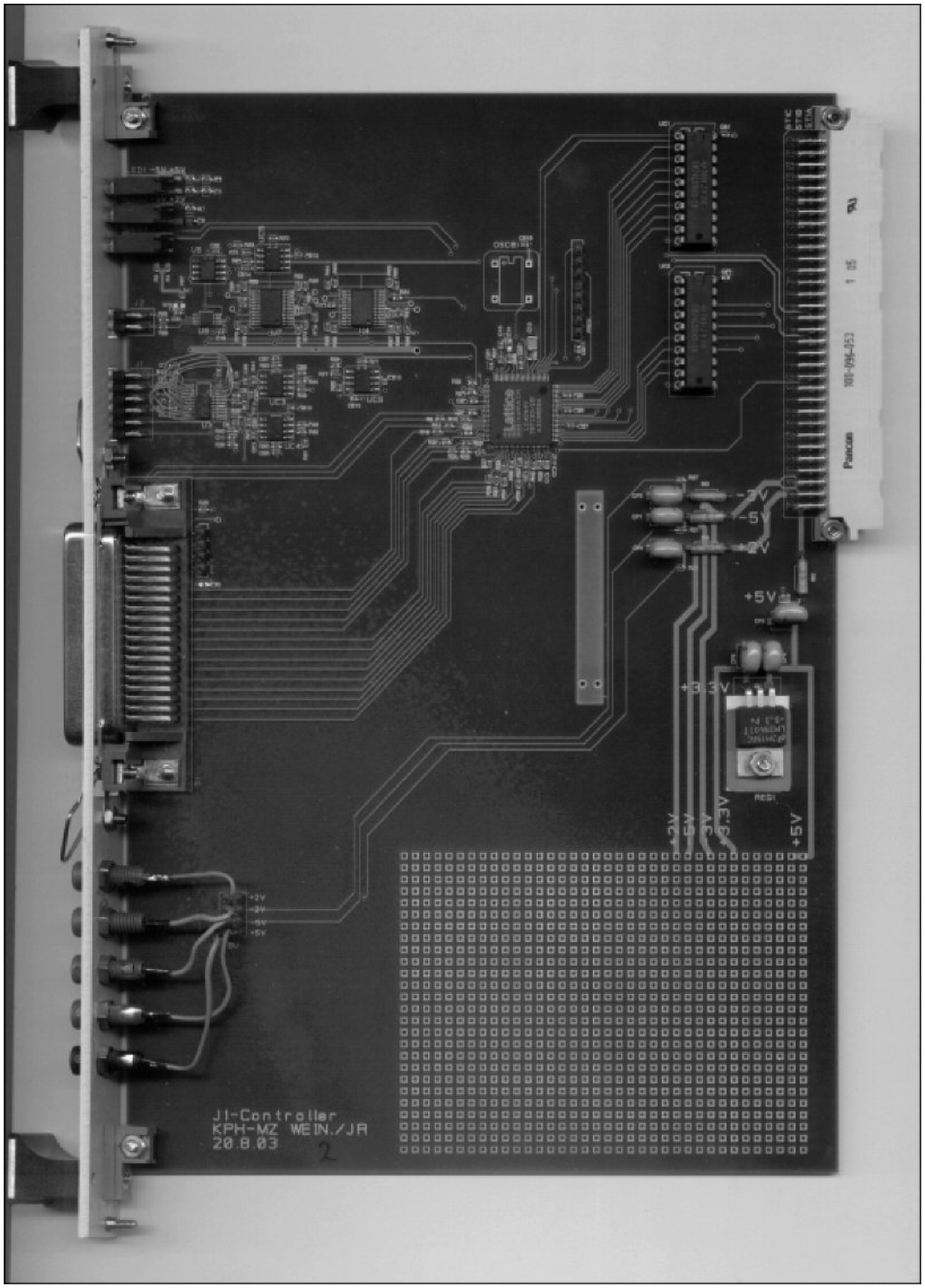}
  \caption{Left: photograph of the double threshold discriminator
  board.  The two LVDS outputs are visible near the bottom of the
  module. The RJ-45 inputs are located at the top left of the module
  and the two Lemo outputs at centre left. Right: photograph of the
  controller board for up to 20 double threshold discriminator
  boards.}
  \label{fig:frontends}
\end{figure}

Two 32-channel discriminator discriminator boards each housing 8 DTD
chips have been prototyped by the electronics workshop of the Institut
f\"ur Kernphysik, see Fig.~\ref{fig:frontends}\,(left) for a
photograph of the DTD board.  The DTD boards have two multiplexed Lemo
analogue outputs for debugging and two LVDS outputs, one to be
connected to TDC modules, and one to trigger modules.  A 32-channel
analogue output board can be attached to the discriminator board for a
complete analysis during the prototyping stage. Up to 20 DTD boards
fit into a VME 6U crate together with a controller board, see
Fig.~\ref{fig:frontends}\,(right) for a photograph. The communication
with a PC is done via parallel port. The trigger will be derived with
GSI VME logic modules. Such a module is equipped with a FPGA, and a
large number of front panel inputs and outputs. The timing is picked
off by TDC CATCH boards. These boards, developed for the COMPASS
collaboration, are equipped with 4 mezzanine cards for a total of 32
channels.  At the heart of the TDC mezzanine cards there are the
so-called ${\cal F}$1 chips, developed by the Faculty of Physics of
the University of Freiburg, Germany, with 8 channels of $\sim 120$\,ps
resolution (LSB) each.

In addition to timing and trigger, an ADC system capable of handling
high channel counts is also needed. For this purpose, another
component of the COMPASS electronics is under investigation, namely
the electronics around the APV-Chip and the GeSiCa data collector
card. This system has been mainly developed by the Faculty of Physics
of the Technical University of Munich, Germany, for the RICH
subdetector of the COMPASS set-up. With 128 input channels, each
performing synchronous analogue sampling at 40\,MHz, the APV chip acts
as an analogue ring-buffer, which on demand multiplexes the
appropriate 128-channel sample and sends it to an attached ADC. The
GeSiCa module provides a similar functionality as the CATCH module,
i.e.\ set-up of the attached front-end electronics, TCS information
processing, data collection, data concentration and data transfer to
PC-based read-out buffer cards via SLink.

\subsection{Prototype Fibre Detector Characterisation at the Electron Beam}
\begin{figure}[htb]
  \includegraphics[width=\columnwidth]{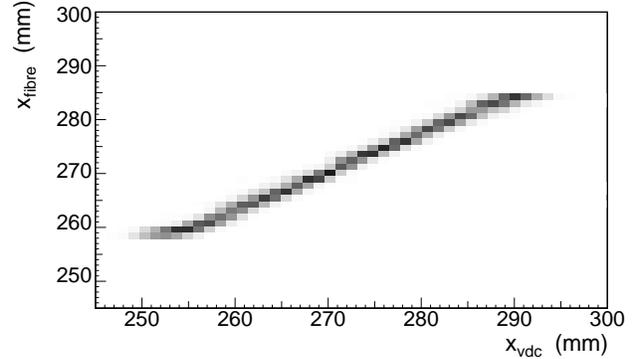}
  \includegraphics[width=0.95\columnwidth]{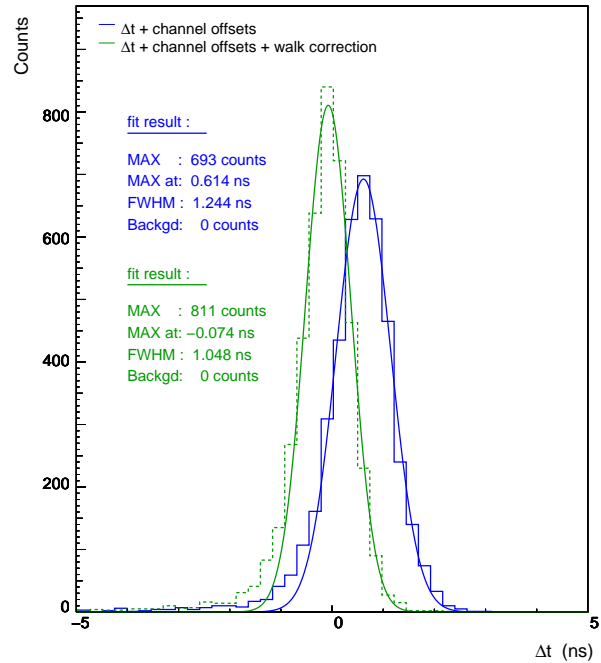}
  \caption{The top panel shows the reconstructed $x$-position of the
  electron projected onto the base coordinate versus the measured
  $x$-position obtained with a simple estimator from the fibre
  detector. The bottom panel shows the time resolution of the fibre
  detector obtained from the coincidence timing with the scintillator
  paddle detectors.}
  \label{fig:beamtime}
\end{figure}
We have employed spectrometer~A of the three spectrometer facility at
MAMI to carry out a characterisation of a 32-channel fibre detector
prototype.  The detector was sandwiched between the drift chambers and
the scintillator paddles of the focal plane detector system. The
arrival time of the electrons was measured in the fibre detector with
respect to the following two overlapping paddles. The electron track
was reconstructed with the position information of the drift chambers
and the electron hit position was extrapolated from the drift chamber
planes to the fibre detector plane.  A simple estimator for the
$x$-position $x= \sum_{i=1}^{N} x(\mbox{fibre}_i)/N + \mbox{offset}$
was used, where $x(\mbox{fibre}_i)$ is the geometrical position of the
$i$th fibre and $N$ the hit multiplicity. That position is compared to
the reconstructed $x$-position of the electron projected onto the
detector base coordinate, see Fig.~\ref{fig:beamtime}\,(top). Small
non-linearities at the edges indicate that better estimators are
needed.  However, position estimators based on weighted averages
suffer from fluctuations in pulse heights.
Fig.~\ref{fig:beamtime}\,(bottom) shows the time resolution obtained
from the coincidence timing with the scintillator paddle detectors
after walk correction of the paddle timing. The FWHM of $\approx
1$\,ns is rather good for the small amount of light from the fibres.

\section{Hypernuclear Gamma-Spectroscopy with PANDA at GSI}
At $\overline{\mbox{\sf P}}${\sf ANDA}\, relatively low momentum
$\Xi^-$ can be produced in $\mathrm{\overline{p}p} \rightarrow \Xi^-
\overline{\Xi}^+$ or $\mathrm{\overline{p}n} \rightarrow \Xi^-
\overline{\Xi}^0$ reactions~\cite{PANDA}. The advantage as
compared to the kaon induced reaction is the fact that the antiproton
is stable and can be retained in a storage ring.  This allows a rather
high luminosity even with very thin primary targets. The associated
$\overline{\Xi}$ will undergo scattering or (in most cases)
annihilation inside the residual nucleus.  Strangeness is conserved in
the strong interaction and the annihilation products contain at least
two anti-kaons that can be used as a tag for the reaction.

Because of the two-step process, spectroscopic studies, based on the
analysis of two-body reactions like in single hypernuclei
reactions~\cite{Agnello2005}, cannot be performed. Spectroscopic
information on double hypernuclei can only be obtained via their
sequential decay, see e.g.\ ~\cite{Aoki1991}. In combination with the
high luminosity at FAIR and with a novel solid-state micro-tracker,
high resolution $\gamma$-ray spectroscopy of double hypernuclei and
$\Omega$ atoms will become possible for the first time.

It is intended to further develop existing HPGe detectors for
$\gamma$-ray spectroscopy in order to use them in the presence of high
particle fluxes and high magnetic fields. This will be achieved by new
read-out schemes and tracking algorithms, which have to be
developed. These studies will include the modelling of detector
response and detailed background studies.

At present, it is foreseen to use a number of Euroball cluster
detectors~\cite{Eberth1996}, partially owned by GSI, and in addition
the use of the Vega detectors~\cite{Gerl1994} of GSI.

Standard electronic readout systems are hardly capable of dealing with
particle rates exceeding 10\,kHz. However, the fully digital
electronic system currently developed for the new generation of Ge
arrays like Vega will -- in connection with fast preamplifiers --
allow the load of background particles at rates higher than 100\,kHz
for each detector element.  Arranging several cluster detectors and
taking into account that most of the produced particles are emitted
into the forward region not covered by the Ge array an interaction
rate of $10^7$\,s$^{-1}$ seems to be manageable.

The Euroball cluster consists of seven large hexagonal tapered Ge
detectors closely packed in a common cryostat. The crystals have a
length of 78\,mm and a diameter of 70\,mm at the cylindrical end. The
super-segmented clover detector Vega has been developed at GSI.  It
consists of four coaxial four-fold segmented Ge detectors of 140\,mm
length and 70\,mm diameter enabling an optimal arrangement with
respect to efficiency and spectrometer response.  Prior to the
measurements the energy resolution of the detectors was deduced with a
$^{60}${Co} source. The Vega detector as well as three of the seven
crystals of the Euroball cluster detector were then set up inside the
ALADiN magnet. A noisy environment was the cause of the
non-perfect resolutions measured without a magnetic field.

\begin{figure}[htb]
  \centerline{\includegraphics[width=\columnwidth]{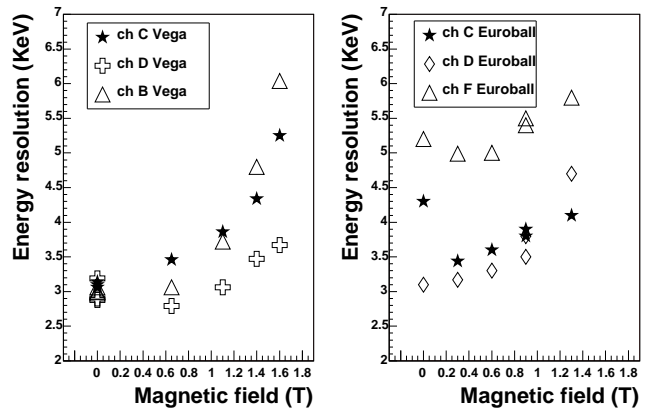}}
  \caption{Measured energy resolution (FWHM) of Euroball cluster and
    Vega detectors in a magnetic field.}
  \label{fig:gam:fig1}
\end{figure}

Fig.~\ref{fig:gam:fig1} shows the energy resolution (FWHM) at
$1.33\,$MeV of three crystals of each detector as a function of the
magnetic field. The energy resolution of one of the Euroball crystals
was found to be worse than the others because of pick-up noise in its
electronic readout, whereas two other crystals of the same detector
showed a similar behaviour in the magnetic field. Their energy
resolution slightly exceeded 0.3\,\% at 1.2\,T which, however, allows
to perform $\gamma$-ray spectroscopy on hypernuclei. A similar
behaviour was found for the Vega detector. For ADC spectra taken
without magnetic field, a Gaussian distribution was chosen for
fitting. In case the magnetic field was non-zero, a fit to a peak is
composed of three components: a Gaussian, a skewed Gaussian, and a
smoothed step function to increase the background on the low-energy
side of the peak.

The ALADiN magnet aperture allows to set the detector only with its
axis in the horizontal plane if the highest magnetic field should be
present in the sensitive part of the detector. Then, the direction of
the magnetic field lines is perpendicular to the detector
axis. However, the energy resolution of the Vega detector was also
measured for other angles. The results obtained indicated no deviation
for $30^\circ$ at a field of $B\sim 0.3\,$T~\cite{Sanchez2005}.

\begin{figure}[htb]
  \centerline{\includegraphics[width=\columnwidth]{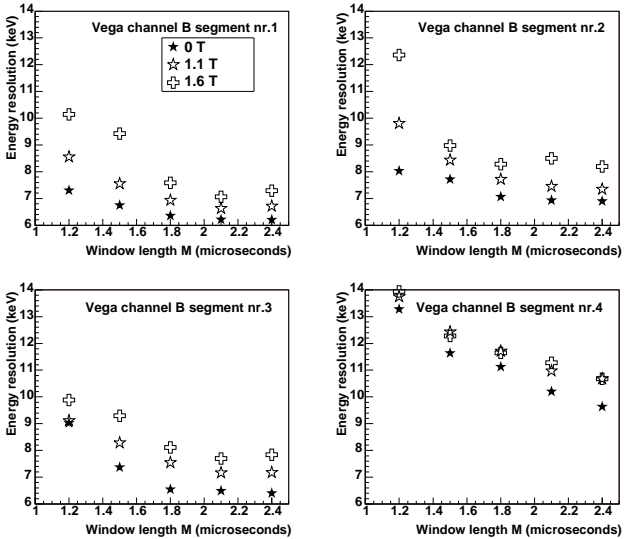}}
  \caption{Measured energy resolution for 4 segments of Vega channel~B
    as a function of the deconvolution window width for
    different values of the magnetic field.}
  \label{fig:vega}
\end{figure}

The measurements were performed over a period of two days without
observing any problems with FETs, vacuum breaks or sparking of the
crystals. After the measurements the original energy resolution was
recovered.

For the measurements of the pulse shape, a SIS3300 8-channel 100\,MHz
12-bit FADC was used. It was directly connected to the preamplifier of
one of the channels of Vega. The scope of this analysis was the
extraction of the energy resolution from the preamplifier signal by
using digital signal processing. The output signal of a charge
integrating preamplifier with continuous discharge consist of a fast
rising step due to the charge collection, followed by an exponential
tail due to the discharge of the capacitors over the resistor. The
exponential tail reduces the final peak height depending on the rise
time of the signal, therefore in order to extract the whole amplitude
of the signal which corresponds to the energy of a gamma ray, the
influence of the preamplifier has to be removed. This is accomplished
by the moving window deconvolution algorithm
(MWD)~\cite{Georgiev1994}.

It consists of two parts: the first part is a deconvolution, which
transforms the continuous discharge preamplifier signal into a
staircase signal. By applying numerical differentiation of such a
signal the moving window deconvolution equation is obtained. It
converts an exponential signal into a step signal of length M, M being
the width of the deconvolution window. Knowing the decay time and the
start time of the signal, the initial amplitude can be determined from
any data point of the decaying signal. The decay time for Vega
detectors is about 50\,$\mu$s.

In Fig.~\ref{fig:vega}, the energy resolution of the 4 segments of the
analysed channel at Vega, extracted by using the method explained
above, is shown as a function of the deconvolution window length M for
different values of the magnetic field. Even though each segment shows
a different character, all segments exhibit a common trend, i.e.\ the
energy resolution improves with longer FADC sampling times.  The pulse
shape is evidently affected at high magnetic field. With these
measurements the investigation of the feasibility of Ge detectors
under high magnetic fields was fully addressed~\cite{Sanchez2005}.

\bigskip 
\begin{acknowledgments}
  We acknowledge financial support from the Bundesministerium f{\"u}r
  Bildung und Forschung (bmb+f) under contract number 06MZ176.  This
  research is part of the EU integrated infrastructure initiative
  HadronPhysics Project under contract number RII3-CT-2004-506078.
\end{acknowledgments}

\bigskip 

\end{document}